# SynthRAD2025 Grand Challenge dataset: generating synthetic CTs for radiotherapy

Adrian Thummerer[1], Erik van der Bijl[2], Arthur Jr Galapon[3], Florian Kamp[4], Mark Savenije[5,6], Christina Muijs[3], Shafak Aluwini[3] , Roel J.H.M. Steenbakkers[3], Stephanie Beuel[4], Martijn PW Intven[5], Johannes A Langendijk[3], Stefan Both[3], Stefanie Corradini[1] , Viktor Rogowski[9,10], Maarten Terpstra[5,6] , Niklas Wahl[11], Christopher Kurz[1], Guillaume Landry[1,7,8], Matteo Maspero[5,6]

[1] Department of Radiation Oncology, LMU University Hospital, LMU Munich, Munich, Germany

[2] Department of Radiation Oncology, Radboud University Medical Center, Nijmegen, The Netherlands;

[3] Department of Radiation Oncology, University Medical Center Groningen, University of Groningen, Groningen, The Netherlands;

[4] Department of Radiation Oncology and Cyberknife Center, University Hospital of Cologne, Cologne, Germany

[5] Department of Radiotherapy, University Medical Center Utrecht, Utrecht, The Netherlands;

[6] Computational Imaging Group for MR Diagnostics & Therapy, University Medical Center Utrecht, Utrecht, The Netherlands;

[7] German Cancer Consortium (DKTK), partner site Munich, a partnership between DKFZ and LMU University Hospital Munich, Germany

[8] Bavarian Cancer Research Center (BZKF), Munich, Germany

[9] Radiation Physics, Department of Hematology, Oncology, and Radiation Physics, Skåne University Hospital, Lund, Sweden

[10] Medical Radiation Physics, Department of Clinical Sciences Lund, Lund University, Lund, Sweden

[11] Deutsches Krebsforschungszentrum (DKFZ), Heidelberg, Germany

# Abstract

## Purpose

Medical imaging is crucial in modern radiotherapy, aiding diagnosis, treatment planning, and monitoring. The development of synthetic imaging techniques, particularly synthetic computed tomography (sCT), continues to attract interest in radiotherapy. The *SynthRAD2025* dataset and the accompanying SynthRAD2025 Grand Challenge aim to stimulate advancements in synthetic CT generation algorithms by providing a platform for comprehensive evaluation and benchmarking of synthetic CT generation algorithms based on cone-beam CTs (CBCT) and magnetic resonance images (MRI).

## Acquisition and validation methods

The dataset comprises 2362 cases, including 890 MRI-CT pairs and 1472 CBCT-CT pairs of head-and-neck, thoracic, and abdominal cancer patients treated at five European university medical centers (UMC Groningen, UMC Utrecht, Radboud UMC (Netherlands), LMU University Hospital Munich, and University Hospital of Cologne (Germany). Images were acquired using a wide range of acquisition protocols and scanners. Pre-processing, including rigid and deformable image registration methods, was performed to ensure high-quality image datasets and alignment between modalities. Extensive quality assurance was performed to validate image consistency and usability.

## Data format and usage notes

All imaging data is provided using the MetaImage (.mha) file format, ensuring compatibility with common medical image processing tools. Metadata, including acquisition parameters and registration details, is available in structured comma-separated value (CSV) files. To ensure dataset integrity, *SynthRAD2025* is split into training (65%), validation (10%), and test (25%) sets. The dataset is accessible through https://doi.org/10.5281/zenodo.14918089 under the *SynthRAD2025* collection.

## Potential applications

This dataset enables benchmarking and development of synthetic imaging techniques for radiotherapy applications. Potential use cases include sCT generation for MRI-only and MR-guided photon and proton radiotherapy, CBCT-based dose calculations, and adaptive radiotherapy workflows. By incorporating data from diverse acquisition settings, *SynthRAD2025* supports the advancement of robust and generalizable image synthesis algorithms for clinical implementation, ultimately promoting personalized cancer care and improving adaptive radiotherapy workflows.

# 1 Introduction

Over the last decade, advancements in image-guided and adaptive radiotherapy have significantly improved treatment outcomes for cancer patients, partially due to the introduction of image-guided (daily) adaptive photon and proton radiotherapy [1]. These approaches rely on accurate imaging to account for anatomical and physiological changes throughout treatment, enabling precise dose delivery to tumor volumes while sparing surrounding healthy tissues [2]. Computed tomography (CT) imaging remains the gold standard for treatment planning, offering the electron density information critical for accurate dose calculations [3]. However, frequent CT imaging is time-consuming and costly, has an additional imaging dose burden for patients, and is usually unavailable directly on radiotherapy delivery machines [4].

To address these challenges, alternative imaging modalities such as cone-beam CT (CBCT) and magnetic resonance imaging (MRI) are increasingly used to replace CT acquisitions during treatment [5,6]. Compact CBCT systems can be easily integrated with treatment machines, providing volumetric patient images, and have become standard for daily pre-treatment patient alignment [7,8]. However, CBCT image quality is usually inferior to diagnostic fan-beam CT quality, mainly due to increased scatter and other CBCT imaging artifacts, which prevent the use of CBCT images for accurate dose calculations [9,10]. Recent advancements in clinically available CBCT hardware and software have enabled direct dose calculations in photon radiotherapy [25], although this approach is not yet widely adopted.

MRI, on the other hand, offers superior soft-tissue contrast and functional imaging capabilities without ionizing radiation. However, direct dose calculations on MRIs are impossible due to the lack of electron density information required for dose calculation algorithms [11]. Still, there is an increasing interest in MR-only radiotherapy workflows [26], and although more technically challenging to realize than compact CBCT systems, MR-Linacs have proven that MRI can be efficiently combined with treatment machines and enable daily MR-guided online adaptive photon radiotherapy [12]. The combination of a treatment machine and an MRI is more challenging for proton therapy due to the interaction between magnetic fields and proton beams; however, research and development are ongoing, and MR-guided proton therapy might become clinically available [13].

The image quality limitations of CBCT and the absence of electron density information in MRI have sparked interest in generating so-called synthetic CTs (sCT) from CBCT and MRI data to enable accurate dose calculations. Beyond generating electron density maps for dose calculations, sCTs have also proven valuable in facilitating organ-at-risk and target volume auto segmentation [23, 24]. Numerous studies highlight artificial intelligence, particularly deep learning, as one of the most promising approaches for synthetic CT generation [5]. However, a lack of public datasets for CBCT and MR-based synthetic CT generation makes a fair and meaningful comparison of deep learning-based synthetic CT algorithms challenging. In 2023, the first edition of the *SynthRAD* challenge, *SynthRAD2023*, addressed this by providing the first large-scale public multi-center dataset to comprehensively compare synthetic CT generation in brain and pelvic patients [13, 14]. The *SynthRAD2025* challenge and dataset build upon the success of the *SynthRAD2023* challenge and provide a public dataset for three additional anatomical locations, head-and-neck, thorax, and abdomen, collected at five European university medical centers. The *SynthRAD2025* dataset aims to support and accelerate research in medical image synthesis for radiotherapy by providing high-quality, curated, and paired CBCT-to-CT and MRI-to-CT datasets. The dataset facilitates the development, validation, and benchmarking of sCT generation algorithms, promoting advancements in radiotherapy and personalized cancer care.

## 2 Acquisition and Validation Methods

### 2.1 Dataset overview

The *SynthRAD2025* dataset is part of the second edition of the SynthRAD deep learning challenge, which focuses on benchmarking MRI- and CBCT-based synthetic CT generation solutions (https://synthrad2025.grand-challenge.org/). Similar to the previous *SynthRAD2023* challenge [14, 15], *SynthRAD2025* is structured into two tasks: Task 1 addresses MRI-to-CT conversion for MR-only and MR-guided photon and proton radiotherapy; Task 2 focuses on CBCT-to-CT translation for daily adaptive radiotherapy workflows. The *SynthRAD2025* challenge dataset provides data for synthetic CT generation in head-and-neck, thoracic, and abdominal cancer patients. Imaging data was collected at radiation oncology departments of five European university medical centers, three from the Netherlands: UMC Groningen, UMC Utrecht, and Radboud UMC, and two from Germany: LMU University Hospital Munich and University Hospital of Cologne.

This study has been independently approved by all centers in accordance with the regulations of their respective institutional review boards or medical ethics committees.

The dataset comprises 2362 cases, where 890 are MRI-CT pairs for task 1 and 1472 are CBCT-CT pairs for task 2. The only inclusion criteria for the *SynthRAD2025* challenge datasets were treatment with some form of external beam radiotherapy (photon- or proton-beam therapy) at one of the data-providing centers and available imaging data from one of the respective anatomical regions. There were no further limitations on age, sex, or tumor characteristics, e.g., type, size, location, and staging. Due to the large dataset size, we have collected a representative sample of patients treated at these radiation oncology departments. Datasets for head-and-neck, thorax, and abdomen subsets were mainly collected based on imaging protocols used in the respective region and institution. Due to this selection, patients from one region were occasionally imaged with the imaging protocol of other regions, e.g., abdomen patients imaged with thorax protocol might be present in the thorax training dataset. However, only patients whose target volumes belonged to the respective anatomical region were selected for the test and validation sets.

In some centers, access to detailed patient characteristics was limited due to ethical considerations of data privacy and anonymization, preventing detailed statistics about age and sex distribution in the dataset. Figure 1 presents exemplary images for each task and anatomical region. For *SynthRAD2025*, the dataset was split into a training, validation, and test set, aiming at a split of 65/10/25%, respectively. However, this may slightly vary depending on data availability per center, task, and anatomy. To ensure the integrity of the *SynthRAD2025* challenge, initially, only the training dataset will be released publicly (details see section 3.2). Table 1 presents the number of cases per set, task, anatomical region, and center. Data-providing centers are abbreviated using the letters A to E. The assigned letter does not align with the order of centers mentioned above. A detailed description of dataset characteristics for each task and anatomy is provided in sections 2.2 and 2.3.

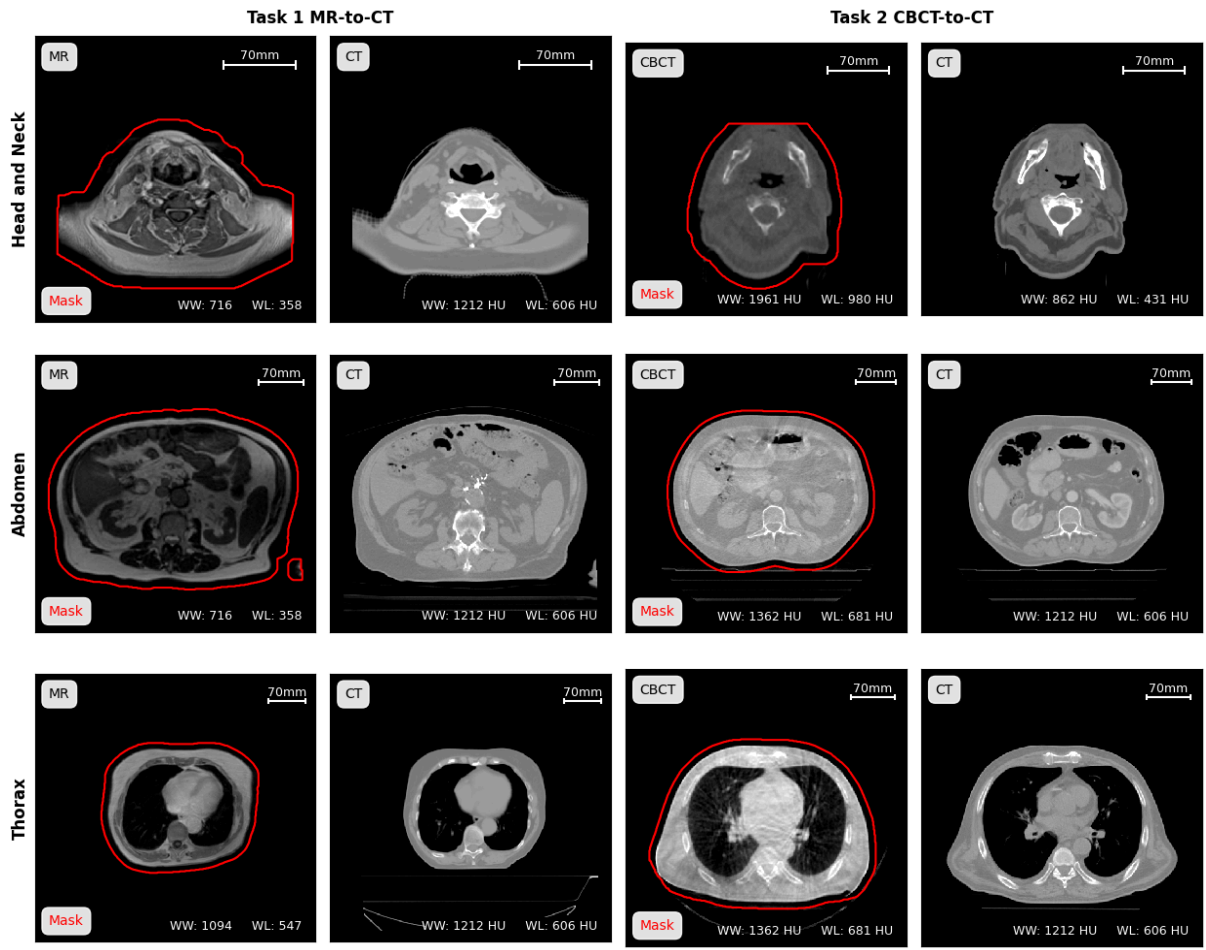

**Figure 1:** *Example images for head-and-neck (top), thorax (middle), and abdomen (bottom) cases of Task 1 (left) and Task 2 (right) of the SynthRAD2025 dataset, with in red the contour of the provided patient outline mask.*

**Table 1:** *The number of cases collected at each center (letter from A to E) for training, validation, and test set for the three anatomical sites: head-and-neck (HN), thorax (TH), and abdomen (AB).*

### Training

| Center | | HN | | | | | | TH | | | | | | AB | | | | |
|---|---|---|---|---|---|---|---|---|---|---|---|---|---|---|---|---|---|---|
| | | A | B | C | D | E | All | A | B | C | D | E | All | A | B | C | D | E | All |
| Task | 1 | 91 | 0 | 65 | 65 | 0 | **221** | 91 | 91 | 0 | 0 | 0 | **182** | 65 | 91 | 19 | 0 | 0 | **175** |
| | 2 | 65 | 65 | 65 | 65 | 65 | **325** | 65 | 65 | 63 | 63 | 65 | **321** | 64 | 65 | 62 | 53 | 65 | **309** |

### Validation

| Center | | HN | | | | | | TH | | | | | | AB | | | | |
|---|---|---|---|---|---|---|---|---|---|---|---|---|---|---|---|---|---|---|
| | | A | B | C | D | E | All | A | B | C | D | E | All | A | B | C | D | E | All |
| Task | 1 | 14 | 0 | 10 | 10 | 0 | **34** | 14 | 14 | 0 | 0 | 0 | **28** | 10 | 14 | 3 | 0 | 0 | **27** |
| | 2 | 10 | 10 | 10 | 10 | 10 | **50** | 10 | 10 | 10 | 10 | 10 | **50** | 10 | 10 | 10 | 8 | 10 | **48** |

**Testing**

|  |  | HN | | | | | | TH | | | | | | AB | | | | | |
|---|---|---|---|---|---|---|---|---|---|---|---|---|---|---|---|---|---|---|---|
| Center |  | A | B | C | D | E | Tot | A | B | C | D | E | Tot. | A | B | C | D | E | Tot |
| Task | 1 | 35 | 0 | 25 | 25 | 0 | **85** | 35 | 35 | 0 | 0 | 0 | **70** | 25 | 35 | 8 | 0 | 0 | **68** |
|  | 2 | 25 | 25 | 25 | 25 | 25 | **125** | 25 | 25 | 25 | 24 | 25 | **124** | 25 | 25 | 25 | 20 | 25 | **120** |

## 2.2 Task 1

Centers A, B, C, and D provided data for task 1, which comprises a variety of image scanners and acquisition protocols. MRIs from centers A, C, and D were acquired for treatment planning, mainly for defining target volumes. MRIs from Center B were acquired on a low-field MR-Linac capable of daily MR imaging and real-time (2D) cine acquisition during treatment. MRIs were acquired with a T1-weighted gradient echo or a balanced steady-state free-precession sequence and collected along with the corresponding planning CTs for all subjects.

### 2.2.1 Head-and-Neck (1HN)

In total, 340 MRI-CT pairs from head-and-neck cancer patients were provided by centers A, C, and D. Image acquisition systems and parameters for MRIs and CTs are presented in Table 2. The data provided by Center C was acquired for diagnostic purposes with a small FOV and different immobilization devices, which makes this dataset specifically challenging for synthetic CT generation. Center A and Center C used similar immobilization and table tops on MRI and CT scanners, providing a similar position on CT and MRI and improving the quality of the pre-processed data. Center A provided 140 cases since no head-and-neck MRI was available from Center B.

*Table 2: Imaging parameters for the head-and-neck MRIs and CTs in Task 1. The dataset is labeled with the prefix "1HN". In parenthesis, the number of cases with a specific parameter is specified, along with the proprietary name of the sequence. A minus sign indicates unavailable or inapplicable parameters.*

| MRI - Head-and-Neck | | | |
|---|---|---|---|
| **Parameter** | **Center A** | **Center C** | **Center D** |
| Manufacturer | Philips | Siemens Healthineers | Siemens Healthineers |
| Model | Ingenia v5.4-7 | Avanto | Skyra (21), Prisma (79) |
| Field Strength [T] | 3 | 1.5/3 | 3 |
| Sequence | T1w spoiled turbo gradient-echo Dixon (TFE) | T1w turbo spin-echo (TSE) | T1w radio-frequency-spoiled gradient echo Dixon (Vibe) |
| Acquisition | 3D | 2D | 3D |
| Contrast | Yes | Yes | Yes |
| Flip angle [ ° ] | 10 | 150/160 | 9 |

| Echo numbers | 2 | 1 | 1 |
|---|---|---|---|
| Echo time [ms] | 1.4, 2.4 | 8.7/11 | 2.46 |
| Repetition time [ms] | 4.4-5.4 | 475-863 | 5.5 |
| Inversion time IR [ms] | - | - | - |
| Number of averages | 1 | 1-3 | 1 |
| Echo train length | 2 (70), 60 (70) | 2/3 | 2 |
| Phase encoding steps | 252 (70),  462 (70) | 150-405 | 278 |
| Bandwidth [Hz/px] | 718-723 | 190-391 | 455 |
| Voxel spacing [mm] | 0.6-1.0 x 0.6-1.0 x 1.1-2.0 | 0.98-1.17 x 0.98-1.17 x 3.0 | 0.9 x 0.9 x 0.9-1.0 |
| Acquisition matrix | 252-462x252-462 x 190-250 | 256-384x 192-288 x 67-320 | 256-264 x 256-264 x 256 |
| Acquisition time [s] | 83 (70), 287 (70) | 82-202 | - |
| **CT** | | | |
| **Parameter** | **Center A** | **Center C** | **Center D** |
| Manufacturer | Philips (100), Siemens (40) | Philips/Siemens | Siemens Healthineers |
| Model | Big Bore (100), Biograph40 (40) | Brilliance Big Bore Biograph40 Somatom Go.Open Pro | SOMATOM Definition AS (81), SOMATOM go.Open Pro (9) |
| kV | 120 | 120 | 120 |
| Tube current [mA] | 128-534 | 60-496 | 76-174 |
| Exposure time [ms] | 614-10000 | 725-1475 | 1000-1250 |
| CTDIvol [mGy] | 15.1-27.5 | 3.8-42.6 | - |
| Rows/Columns | 512 | 512 | 512 |
| Pixel spacing [mm, mm] | 0.9-1.3 x 0.9-1.3 | 0.98-1.37 x 0.98-1.37 | 0.98 x 0.98 |
| Slice thickness [mm] | 2 | 3 | 2 |
| Reconstruction Diameter [mm] | 451-700 | 500-700 | 500 |

## 2.2.2 Thorax (1TH)

Only two of the five data-providing institutes had suitable thoracic MRIs. Two hundred eighty images were collected, with equal contributions from centers A and B. The respective image acquisition parameters are listed in Table 3.

*Table 3:* *Imaging parameters for the thorax MRIs and CTs in Task 1. The dataset is labeled with the prefix "1TH". In parenthesis, the number of cases with a specific parameter is specified, along with the proprietary name of the sequence. A minus sign indicates unavailable or inapplicable parameters.*

| MRI - Thorax | | |
|---|---|---|
| **Parameter** | **Center A** | **Center B** |
| Manufacturer | Philips | ViewRay |
| Model | Ingenia v5.1-7 | MRidian |
| Field Strength [T] | 1.5 | 0.35 |
| Sequence | T1w spoiled gradient-echo Dixon (70, TFE)/ T1w radial fat-suppressed gradient echo  (70, VANE) | balanced steady-state free-precession sequence (bSSFP, TrueFISP) |
| Acquisition | 3D | 2D |
| Contrast | No | No |
| Flip angle [ ° ] | 10-12 | 60 |
| Echo numbers | 1 | 1 |
| Echo time [ms] | 2.3-4.7 | 1.27 - 1.62 |
| Repetition time [ms] | 5.5-7.4 | 3.0 - 3.8 |
| Inversion time IR [ms] | - | - |
| Number of averages | 1-4 | 1 |
| Echo train length | 400 (70), 105-120 (70) | - |
| Phase encoding steps | 320-460 | 175-232 |
| Bandwidth [Hz/px] | 718-723 | 385-604 |
| Voxel spacing [mm] | 0.9-1.3 x 0.9x1.3 x 2.5-3.0 | 1.5-1.63 x 1.5-1.63 x 1.5-3.0 |
| Acquisition matrix | 400-460x400-460 (70) 280-300x280-300 (70) | 200-310x234-360 |
| Acquisition time [s] | 188-340 | 17-25 |
| CT | | |
| **Parameter** | **Center A** | **Center B** |
| Manufacturer | Philips (136), Siemens (4) | Toshiba |
| Model | Big Bore (136), | Aquilion/LB |

|  | Biograph40 (4) |  |
|---|---|---|
| kV | 120 | 120 |
| Tube current [mA] | 31-271 | 40-417 |
| Exposure time [ms] | 615-10829 | 500-750 |
| CTDIvol [mGy] | 2.7-35.5 | - |
| Rows/Columns | 512 | 512 |
| Pixel spacing [mm] | 0.9-1.4 x 0.9-1.4 | [0.82-1.52, 0.82-1.52] |
| Slice thickness [mm] | 2 (30) - 3 (110) | 3 |
| Reconstruction Diameter [mm] | 461-700 | 500-700 |

### 2.2.3 Abdomen (1AB)

Centers A, B, and C provided 270 abdominal MRI-CT pairs in total, while Center C could only provide 30 cases. Center A provided 95 cases, and Center B compensated for the low number of Center C cases, which were 140. MRI and CT acquisition parameters are described in Table 4.

*Table 4:* Imaging parameters for the abdominal MRIs and CTs in Task 1. The dataset is labeled with the prefix "1AB". In parenthesis, the number of cases with a specific parameter is specified, along with the proprietary name of the sequence. A minus sign indicates unavailable or inapplicable parameters.

| MRI - Abdomen | | | |
|---|---|---|---|
| **Parameter** | **Center A** | **Center B** | **Center C** |
| Manufacturer | Philips | ViewRay | Philips/Siemens |
| Model | Ingenia v5.1-7 | MRidian | Marlin(21);Avanto |
| Field Strength [T] | 1.5 | 0.35 | 1.5(21);1.5/3.0(9) |
| Sequence | T1w spoiled gradient-echo Dixon (50, TFE)/ T1w radial fat-suppressed gradient echo (50, VANE) | Balanced steady-state free-precession sequence (bSSFP, TrueFISP) | SE(21);SE/GR |
| Acquisition | 3D | 2D | 3D(21);2D/3D |
| Contrast | No (70), Yes (30) | No | No |
| Flip angle [ ° ] | 8-12 | 60 | 90(21);49-180 |
| Echo numbers | 1 | 1 | 1 |
| Echo time [ms] | 2.3-4.6 | 1.27 - 1.62 | 124(21);1.9-205 |
| Repetition time [ms] | 5.4-6.8 | 3.0 - 3.8 | 1300(21);480-2040 |

| Inversion time IR [ms] | - | - | - |
|---|---|---|---|
| Number of averages | 1-5 | 1 | 2(21);1-3 |
| Echo train length | 58-200 | - | 100(21);1-34 |
| Phase encoding steps | 352-412 | 175-232 | 347(21) |
| Bandwidth [Hz/px] | 433-725 | 385-604 | 820(21) |
| Voxel spacing [mm] | 0.9-1.3 x 0.9-1.3 x 2.2-7.0 | 1.5-1.6 x 1.5-1.6 x 3.0 | 0.64x0.64x2(21) |
| Acquisition matrix | 336-412 x 336-412 130 -273 | 200-310 x 234-360 | 347x347x110(21) |
| Acquisition time [s] | 123-332 | 17-175 | 84-154(21) |
| **CT** | | | |
| **Parameter** | **Center A** | **Center B** | **Center C** |
| Manufacturer | Philips | Toshiba | Philips/Siemens |
| Model | Big Bore | Aquilion/LB | Brilliance Big Bore/Somatom Go.Open Pro |
| kV | 90-120 | 120 | 120 |
| Tube current [mA] | 47-305 | 40-420 | 68-429 |
| Exposure time [ms] | 614-10091 | 500-750 | 421-11912 |
| CTDIvol [mGy] | 15.1-79.6 | - | 5.6-141 |
| Rows | 512 | 512 | 512 |
| Pixel spacing [mm] | 0.9-1.4 x 0.9-1.4 | 0.6-1.4 x 0.6-1.4 | 0.98-1.17x0.98-1.17 |
| Slice thickness [mm] | 2 (8) - 3 (132) | 3 | 2-3 |
| Reconstruction Diameter [mm] | 500-700 | 320-700 | 500-700 |

## 2.3 Task 2

Thanks to the widespread use of image-guided radiotherapy based on CBCT in clinical practice, CBCTs were available in all five participating centers for all anatomical regions, leading to 1496 CBCT-CT pairs. Data was acquired on three different treatment machines/CBCT systems, representing many clinically used CBCT scanners and acquisition protocols.

### 2.3.1 Head-and-Neck (2HN)

The Head-and-Neck CBCT subset features datasets from Elekta (center A, B, C, D) and Varian (center E) linear accelerators (linac) and an IBA proton therapy machine (center D). Table 5 lists the parameters of CBCT and CT image acquisition.

*Table 5:* *Imaging parameters for the head-and-neck CBCTs and CTs in Task 2. The dataset is labeled with the prefix "2HN". In parenthesis, the number of cases with a specific parameter. A minus sign indicates unavailable or inapplicable parameters.*

| CBCT - Head and Neck | | | | | |
|---|---|---|---|---|---|
| **Parameter** | **Center A** | **Center B** | **Center C** | **Center D** | **Center E** |
| Manufacturer | Elekta | Elekta | Elekta | IBA (97), Elekta (3) | Varian |
| Model | XVI v5.x | XVI v5.52 | XVI v5.x | Proteus P+, XVI v5.x | TrueBeam OBI |
| kVp | 100-120 | 100 | 120 | 100 | 100-125 |
| Tube current [mA] | 12-20 | 10 | 10-20 | 160 | 11-20 |
| Exposure Time [ms] | 10-32 | 10 | 22 | 3225 | 7500-18060 |
| Rows/Columns | 270 | 270 | 270 | 270-512 x 270-512 | 512x512 |
| Pixel spacing [mm] | 1 x 1 | 1 x 1 | 1x1 | 0.5-1 x 0.5-1 | 0.5-0.9 x 0.5-0.9 |
| Slice thickness [mm] | 1 | 1 | 1 | 2-2.5 | 2 |
| Reconstruction Diameter [mm] | 270 | 270 | N/A | 260 | 262 - 465 |
| CT | | | | | |
| **Parameter** | **Center A** | **Center B** | **Center C** | **Center D** | **Center E** |
| Manufacturer | Philips, Siemens | Toshiba | Philips, Siemens | Siemens Healthineers | TOSHIBA, Siemens |
| Model | Big Bore (90), Biograph40 (10) | Aquilion/LB | Brilliance Big Bore(93), Biograph40 (7) | SOMATOM Confidence(??)/Definition(??) | Aquilion/LB, Biograph128 |
| kV | 120 | 120 | 120 | 120 | 120 |
| Tube current [mA] | 55-531 | 40-300 | 56-444 | 19-219 | 25-200 |
| Exposure time [ms] | 615-1000 | 500-1000 | 922-1457 | 1000 | 500-1000 |
| CTDIvol [mGy] | 15.1-27.5 | 7.3-117.8 | 10.1-38.6 | - | - |
| Rows/Columns | 512 | 512 | 512 | 512 | 512 |

| Pixel spacing [mm] | 0.7-1.4 x 0.7-1.4 | 1.1-1.4 x 1.1-1.4 | 1-1.2 x 1-1.2 | 1-1.6 x 1-1.6 | 1-1.5 x 1-1.5 |
|---|---|---|---|---|---|
| Slice thickness [mm] | 2 - 3 | 1 - 3 | 2-3 | 2 | 3-5 |
| Reconstruction Diameter [mm] | 444-700 | 550-700 | 350-700 | 500-800 | 531-780 |

### 2.3.2 Thorax (2TH)

In the thoracic region, CBCTs were acquired with various treatment machines: Centers A, B, C, and D used an Elekta linac, Center D an IBA proton cyclotron, and Center E a Varian linac. Table 6 lists detailed image acquisition parameters.

*Table 6: Imaging parameters for the thorax CBCTs and CTs in Task 2. The dataset is labeled with the prefix "2TH". In parenthesis, the number of cases with a specific parameter. A minus sign indicates unavailable or inapplicable parameters.*

| CBCT - Thorax | | | | | |
|---|---|---|---|---|---|
| **Parameter** | **Center A** | **Center B** | **Center C** | **Center D** | **Center E** |
| Manufacturer | Elekta | Elekta | Elekta | IBA (90), Elekta (7) | Varian |
| Model | XVI v5.x | XVI v5.x | XVI v5.x | Proteus P+, XVI v5.x | TrueBeam OBI |
| kV | 100-120 | 120 | 120 | 110-120 | 100-125 |
| Tube current [mA] | 20-40 | 40 | 10-40 | 16-320 | 13-80 |
| Exposure time [ms] | 10-40 | 40 | 16-40 | 10-5900 | 1710-18120 |
| Rows/Columns | 270 | 410 | 135-410 | 270-768 x 270-768 | 512 x 512 |
| Pixel spacing [mm] | 1 x 1 | 1 x 1 | 1-2 x 1-2 | 0.46-1 x 0.46-1 | 0.5-0.9 x 0.5-0.9 |
| Slice thickness [mm] | 1 | 1 | 1-2 | 2-2.5 | 2 |
| Reconstruction Diameter [mm] | 270 | 410 | N/A | 350-500 | 262 - 465 |
| CT | | | | | |
| **Parameter** | **Center A** | **Center B** | **Center C** | **Center D** | **Center E** |
| Manufacturer | Philips (98), Siemens (2) | Toshiba | Philips/Siemens | Siemens | Toshiba, Siemens |
| Model | Big Bore (98), | Aquilion/LB | Brilliance Big Bore(90)/So | SOMATOM Confidence, | Aquilion/LB, Biograph128 |

|  | Biograph (2) |  | matom Go.Open Pro(9)/Biograph40(1) | Definition, go.Open Pro |  |
|---|---|---|---|---|---|
| kV | 100 (2)-120 (98) | 120 | 120 | 120 | 120 |
| Tube current [mA] | 35-295 | 40-440 | 33-502 | 21-243 | 17-440 |
| Exposure Time [ms] | 500-10837 | 500-800 | 437-11914 | 500-6222 | 500-1000 |
| CTDIvol [mGy] | 3.1-35.5 | - | 2.3-40.6 | - | - |
| Rows/Columns | 512 | 512 | 512/1024 | 512 | 512 |
| Pixel spacing [mm] | 0.8 x 1 | 1.0-1.4 x 1.0-1.4 | 0.5-1.5 x 0.5-1.5 | 0.8-1.6 x 0.8-1.6 | 1-1.5 x 1-1.5 |
| Slice thickness [mm] | 2(6) - 3(94) | 2.5-3.0 | 2-3 | 2 | 3 - 5 |
| Reconstruction Diameter [mm] | 486-700 | 500-700 | 500-750 | 398-800 | 500-780 |

### 2.3.3 Abdomen (2AB)

The collected abdomen CBCTs were predominantly acquired on linear accelerators (Elekta and Varian), and only a minimal number of abdominal cancer patients were treated with proton therapy (IBA) in center D. Acquisition parameters of CBCTs and corresponding CTs are presented in Table 7.

*Table 7: Imaging parameters for the abdomen CBCTs and CTs in Task 2. The dataset is labeled with the prefix "2AB". In parenthesis, the number of cases with a specific parameter. A minus sign indicates unavailable or inapplicable parameters.*

| CBCT - Abdomen | | | | | |
|---|---|---|---|---|---|
| **Parameter** | **Center A** | **Center B** | **Center C** | **Center D** | **Center E** |
| Manufacturer | Elekta | Elekta | Elekta | Elekta (70), IBA (11) | Varian |
| Model | XVI 5.x | XVI | XVI v5.x | XVI v5.x , Proteus P+ | TrueBeam OBI |
| kV | 100-120 | 120 | 120 | 120-125 | 125-140 |
| Tube current [mA] | 20-64 | 40 | 10-40 | 16-320 | 15-99 |
| Exposure time [ms] | 10-40 | 40 | 20-40 | 10-5900 | 750-18280 |
| Rows/Columns | 270-410 | 410 | 270-410 | 270-768 | 512x512 |

| Pixel spacing [mm] | 1 x 1 | 1 x 1 | 1x1 | 0.65-1 x 0.65-1 | 0.5-0.9 x 0.5-0.9 |
|---|---|---|---|---|---|
| Slice thickness [mm] | 1 | 1 | 1 | 2-2.5 | 2 |
| Reconstruction Diameter [mm] | 270-410 | 410 | N/A | 270-500 | 262-465 |

| CT | | | | | |
|---|---|---|---|---|---|
| **Parameter** | **Center A** | **Center B** | **Center C** | **Center D** | **Center E** |
| Manufacturer | Philips (93), Siemens (7) | Toshiba, GE | Philips(93)/ Siemens(7) | Siemens Healthineers, GE Medical | Toshiba |
| Model | Big Bore (93), Biograph (7) | Aquilion/LB, Discovery 690 | Brilliance Big Bore Biograph40 (3) Somatom Go.Open Pro (4) | SOMATOM Confidence, Definition, go.Open Pro, Optima CT580 | Aquilion/LB |
| kV | 90 (4), 100 (30), 120 (66) | 120 | 120 | 80-140 | 120 |
| Tube current [mA] | 40-419 | 40-440 | 76-500 | 30-419 | 17-440 |
| Exposure Time [ms] | 500-10837 | 500-800 | 500-11908 | 437-8720 | 500-1500 |
| CTDIvol [mGy] | 3.1-71 | 3.5-108.4 | 10-74.8 | - | - |
| Rows/Columns | 512 | 512 | 512 | 512 | 512 |
| Pixel spacing [mm] | 0.7 x 1.4 | 0.78 - 1.37 x 0.78 - 1.37 | 0.98-1.47x0.98-1.47x | 0.77-1.56 x 0.77-1.56 | 1-1.52 x 1-1.52 |
| Slice thickness [mm] | 2(33) - 3(67) | 2-3 | 3 | 2-2.5 | 1-5 |
| Reconstruction Diameter [mm] | 452-700 | 381-700 | 500-700 | 395-800 | 530-780 |

## 2.4 Pre-processing

The preprocessing workflow aimed to harmonize image parameters (e.g., voxel spacing, orientation), anonymize images, reduce file size, generate patient outlines, and prepare datasets for synthetic CT evaluation. The preprocessing code is publicly available at https://github.com/SynthRAD2025/preprocessing.

Each participating center exported MRIs, CBCTs, and corresponding planning CTs from their clinical databases. For a subset of patients included in the testing phase of *SynthRAD2025*, radiotherapy treatment planning structures were exported and preprocessed. These structures are not included in the released dataset. Following the raw data export, the preprocessing pipeline involved the following key steps.

### 2.4.1 Rigid registration

MRIs and CBCTs were rigidly registered to their corresponding planning CTs using the Elastix registration framework [16]. Parameter files were tested and optimized for each task and anatomical region, and the final parameter files used in preprocessing are included in the public repository.

### 2.4.2 Defacing

Images with visible facial structures were defaced to ensure patient anonymity using an automated algorithm developed for the *SynthRAD2025* dataset. This algorithm utilizes TotalSegmentator (version 2.3.0) [17], a deep learning-based CT auto-segmentation model, to segment the skull and brain. Using these structures, the facial region was identified on the central sagittal slice of the brain mask by extracting the most anterior voxel of the brain (indicated by the yellow marker in Figure 2) and generating a bounding box around the skull mask. The anterior-inferior corner of this bounding box was selected as the lower boundary of the face (see blue marker in Figure 2). The facial region was then defined as all voxels to the left of the line connecting these two points and was overwritten with the background intensity value (-1024 for CT and 0 for MRI). The algorithm demonstrated strong robustness against patient positioning and orientation changes and did not require manual corrections. Figure 2 provides a visualization of the defacing process.

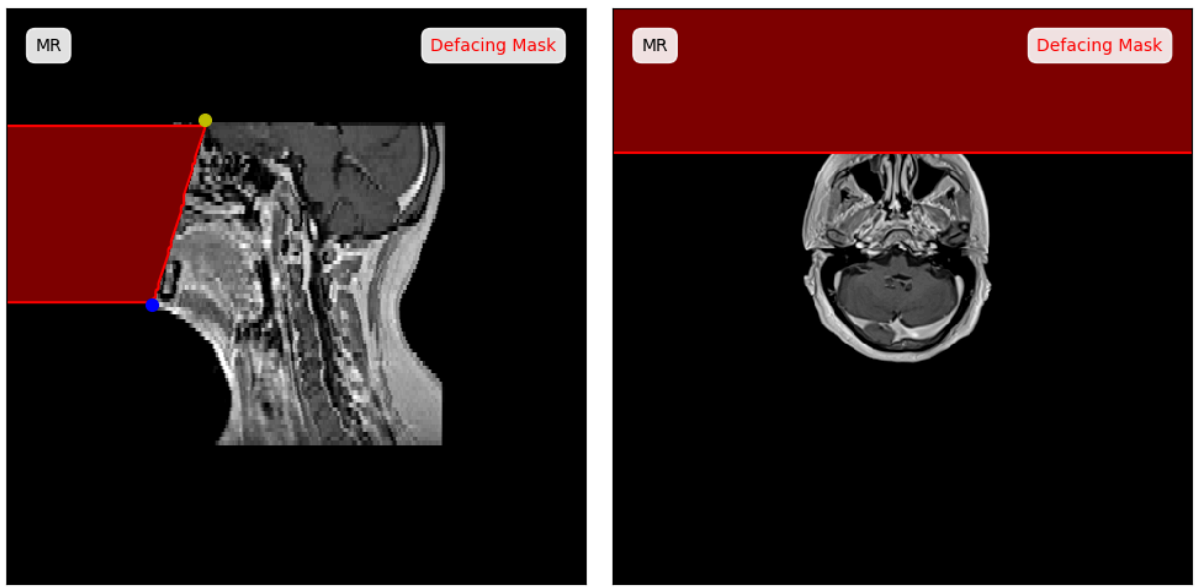

*Figure 2: Example of the automatic defacing algorithm for the SynthRAD2025 dataset. The red area indicates the region overwritten with background values during the defacing. The yellow and blue markers indicate the points defining the defacing mask and were derived from the auto-segmented brain (yellow) and mandible (blue) structures on the corresponding CT.*

### 2.4.3 Resampling

To standardize image resolution across the MRI, CBCT, and CT datasets, all images were resampled with a consistent voxel spacing of 1 × 1 × 3 mm.

### 2.4.4 Outline Segmentation

For each patient case, an outline mask was generated on the MR/CBCT image to define the volume used for evaluation and metric calculations during the validation and testing stage of the *SynthRAD2025* challenge. This mask was created automatically using histogram-based thresholding, followed by morphological erosion and dilation operations. The threshold value varied between centers' anatomical regions and had to be manually tuned for some patients. The final mask was dilated further to include surrounding air, ensuring that synthetic CT models accurately reconstruct the patient outline rather than relying on the mask itself. The automated process can result in minor inaccuracies and variable dilation margins for some patients. Given the large dataset size, manual corrections of masks were not feasible.

### 2.4.5 Cropping
To minimize file and dataset size, patient images were cropped to a 10 pixels-extended bounding box of the patient outline mask (described in Section 2.4.4).

### 2.4.6 File conversion
Images were compressed and saved in the MetaImage format for the final datasets with the ".mha" extension. Pixel data was stored in INT16 to further reduce the file size.

### 2.4.6 Deforming CT to MRI/CBCT
To evaluate image similarity and dose calculation accuracy, CTs were deformed to match the anatomy of the input MR or CBCT and generate ground truth CTs. This step reduced anatomical differences between synthetic and ground-truth CTs. Deformable image registration was performed using the Elastix framework [16], and parameter files are publicly available in the source code. To avoid bias towards paired training approaches, deformed CTs are not provided for the training dataset and will only be released as part of the validation and testing dataset for *SynthRAD2025*.

## 2.5 Data validation
The *SynthRAD2025* dataset was designed to provide a representative sample of radiotherapy patients sourced from multiple international radiation oncology departments. Inclusion criteria were intentionally broad, including patients with even image artifacts or implants, provided the images were deemed suitable for synthetic CT generation. The dataset was validated by focusing on image and preprocessing quality checks, with particular emphasis on the accuracy of the defacing algorithm to ensure patient anonymity and prevent re-identification. Therefore, all datasets underwent visual checks by the respective institutions to ensure proper removal of facial features.

Further quality assurance involved generating overview images containing central axial, sagittal, and coronal slices from CBCT/MRI, CT, and the patient outline mask. These overviews also included overlaid CBCTs/MRIs and CTs to assess registration accuracy visually. It is important to note that these overviews are unsuitable for image intensity quantification due to inherent differences in intensity and contrast between imaging modalities. However, they are distributed as part of the *SynthRAD2025* dataset (Section 3.1). The large dataset volume limited the quality checks to three planes per image and patient.

Available Image acquisition parameters were extracted as is from the original Dicom files and are provided for each dataset in .xlsx files. The selection of patient cases for training, validation, and test sets was guided by the availability of organs-at-risk (OAR) and target structures and the accuracy of deformable image registration, which was visually assessed for all patients.

During the visual control, the following observations were made: 1) In some cases, the position of the arms varied between MR/CBCT and CT acquisitions. 2) Image artifacts, such as those caused by metal implants, were present in a limited number of cases. 3) Depending on the definition of anatomical regions and imaging protocols in each center, some thoracic cases are included in the abdominal dataset and vice versa. 4) Variations among patients affected the automatic thresholding process for the definition of the body mask, resulting in the possible inclusion of couch structures or the exclusion of lung regions. As a result of the automatic thresholding and the varying thresholds used, the final dilation margins around the patient outline are different among patients and dataset. 5) The 1HN subset of center C included MRIs with a limited field of view, making rigid and deformable registration particularly challenging. These cases may be challenging for synthetic CT (sCT) generation. 6) Patient outline masks in subsets 1AB and 1TH of center B were cropped in the inferior-superior direction due to varying MRI intensities and frequent artifacts at the edge of the FOV. In some cases, the cropped mask still includes artifacts, or the cropping might remove regular slices.

## 3 Data Format and Usage Notes

### 3.1 Data structure and file formats

Figure 3 presents the directory structure of the *SynthRAD2025* training dataset. Similar to the *SynthRAD2023* challenge, the dataset is split into the two investigated tasks: Task 1 directory contains all  MRI cases, and Task 2 contains all CBCT cases. Within each task, individual folders exist for each anatomical region: head-and-neck (HN), thorax (TH), and abdomen (AB). Within these anatomy directories, individual folders exist per case. Each case was assigned a unique seven-letter alphanumeric code: a task identifier (1 or 2), a region identifier (HN, TH, or AB), a center identifier, and a three-digit patient ID. Each patient folder contains the input image (mr.mha or cbct.mha), the corresponding CT (ct.mha), and the patient outline mask (mask.mha). The patient overview images and a spreadsheet with image acquisition parameters are provided in an overview directory in each region folder. The deformed CT (ct_def.mha) will also be included in the patient directories for the validation and testing datasets.

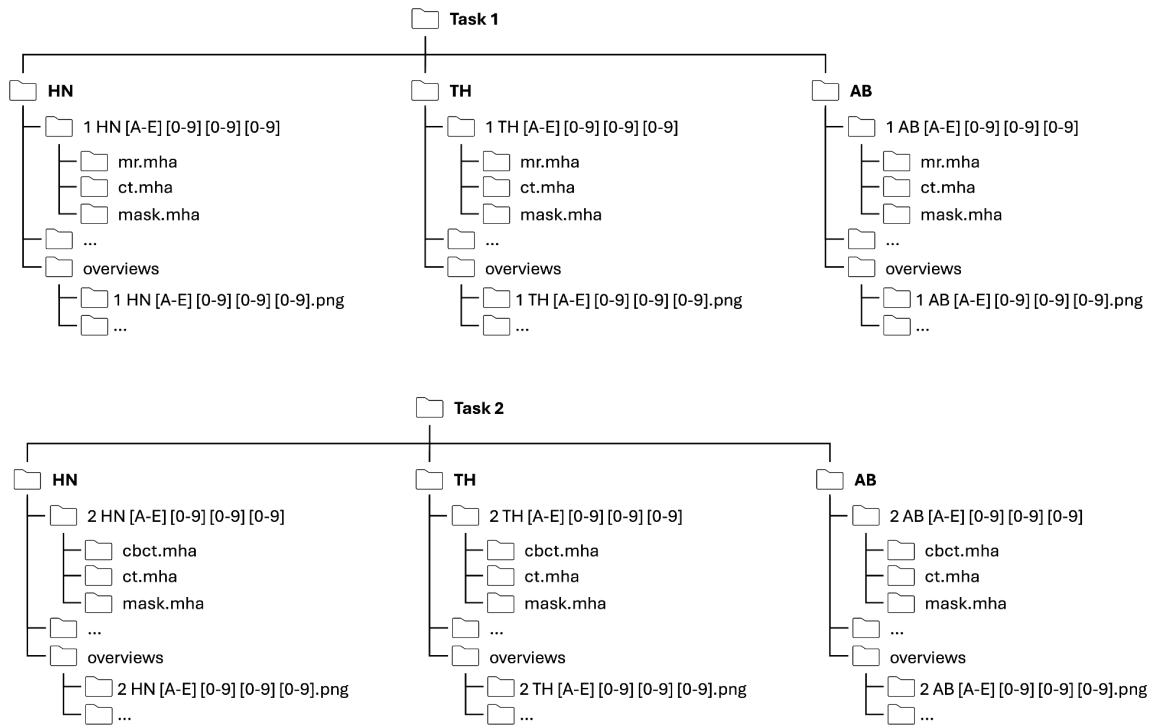

*Figure 4: Folder structure of the SynthRAD2025 training dataset, split based on task and anatomy.*

The dataset is provided under two different licenses. Data from centers A, B, C, and E is provided under a CC-BY-NC 4.0 International License (creativecommons.org/licenses/by-nc/4.0/). Table 8 provides an overview of the release dates and the files included in the SynthRAD2025 training, validation, and test set.

Data from center D is provided with a limited license that allows the use of the data solely for the challenge duration and is only valid as long as the challenge is active. Afterwards, the data download will be deactivated, and the data must be deleted. After requesting participation in the challenge on the SynthRAD2025 website, participants can access the download link for center D at https://synthrad2025.grand-challenge.org/data/.

*Table 8:  Included files, release dates, and links to the dataset download for training, validation, and test sets.*

| Subset | Files | Release Date | Link |
|---|---|---|---|
| **Training** | Input, CT, Mask | 01.03.2025 | https://doi.org/10.5281/zenodo.14918214 |
| **Validation** | Input, Mask | 01.06.2025 | https://doi.org/10.5281/zenodo.14918505 |
| **Validation** | CT, Deformed CT | 01.03.2030 | https://doi.org/10.5281/zenodo.14918606 |
| **Testing** | Input, CT, Deformed CT, Mask | 01.03.2030 | https://doi.org/10.5281/zenodo.14918723 |

## 3.2 Usage notes

All images are compressed files with .mha extension that can be read, written, and modified using the ITK open-source framework (https://itk.org) [18]. For various programming languages, such as Python, R, Java, C++, etc., SimpleITK [19] provides a simplified interface to ITK (https://simpleitk.org/). The pre-processing scripts in the SynthRAD repository contain examples of basic image processing using SimpleITK. Several graphical user interface-based applications exist for viewing .mha images, including 3DSlicer (https://www.slicer.org/), itksnap (http://www.itksnap.org/), and vv (https://github.com/open-vv/vv).

# 4 Discussion

The *SynthRAD2025* dataset is a comprehensive resource designed to advance research on synthetic CT (sCT) generation for radiotherapy. It includes detailed imaging data that supports developing, refining, and benchmarking algorithms for CT image synthesis. This initiative addresses a critical gap in the field by providing a large-scale, multi-center, multi-vendor, and publicly available dataset. It enables researchers to develop, validate, and benchmark sCT generation algorithms for MRI-to-CT and CBCT-to-CT tasks. Below, we discuss this work's implications, strengths, limitations, and future directions.

## 4.1 Implications for adaptive radiotherapy

The *SynthRAD2025* dataset has the potential to accelerate advancements in adaptive radiotherapy by addressing two key challenges: the limitations of CBCT image quality and the lack of electron density information in MRI. By facilitating the development of robust sCT generation algorithms, this dataset can improve the accuracy of dose calculations in MR-guided and CBCT-guided adaptive radiotherapy workflows. This is particularly relevant for proton therapy, where precise dose delivery is critical due to the sensitivity of proton beams to anatomical changes [20]. Including multiple anatomical regions (head-and-neck, thorax, and abdomen) and data from various international institutions ensures that the dataset is representative of diverse clinical scenarios, making it a valuable resource to develop and benchmark robust algorithms for photon and proton radiotherapy.

## 4.2 Strengths of the dataset

The multi-modal and multi-center *SynthRAD2025* dataset includes data from five European university medical centers, ensuring diversity in imaging protocols, patient populations, and treatment machines. This multi-center approach enhances the generalizability of algorithms developed using the dataset, potentially reducing the hurdle of translating research results into clinical practice. With almost 2400 cases, the *SynthRAD2025* dataset is the most extensive curated and publicly available dataset specifically targeted at sCT generation in radiotherapy. The rigorous automatic preprocessing pipeline followed by manual quality control ensures high-quality and standardized data, including rigid registration, defacing, resampling, and outline segmentation. The separation into two separate tasks, addressing MRI-to-CT and CBCT-to-CT conversion, allows researchers to focus on specific challenges associated with each modality, such as the lack of electron density in MRI or the artifacts in CBCT, individually. Furthermore, it enables the investigation of which deep learning model suits each task best. The dataset is publicly available under open licenses, promoting transparency and reproducibility in research. The preprocessing code and parameter files are also publicly available, allowing the reproduction of the research and the reuse of private data with similar characteristics.

### 4.3 Limitations and challenges

While the multi-center design is a strength, it also introduces variability in imaging protocols, such as differences in MRI sequences, CBCT acquisition parameters, and CT reconstruction methods. Furthermore, even within the centers, imaging protocols and scanners often vary, limiting the number of datasets with homogenous image characteristics. This heterogeneity makes it challenging to develop universally applicable sCT generation algorithms.

Due to ethical considerations and data privacy concerns that vary among countries and centers, detailed patient characteristics, e.g., age, sex, tumor type, and staging, are not uniformly available across the dataset. This limits the ability to perform subgroup analyses or evaluate algorithm performance in specific patient populations over the whole dataset. Whenever possible, patient characteristics were included in the metadata files.

Although the preprocessing pipeline was designed to harmonize the data, some steps, such as resampling, defacing, and outline segmentation, may partially deteriorate the data quality. For example, while robust, the automated defacing algorithm may occasionally remove non-facial structures, and the patient outline masks may have variable dilation margins or include structures outside the patient, e.g., the treatment couch.

Deformed CTs are not provided for the training dataset in the *SynthRAD2025* challenge to avoid bias toward paired deep learning training approaches. While this ensures a fair evaluation of synthetic CT algorithms, it may limit the ability to train models that rely on deformable image registration and require extra steps from the dataset user to perform deformable registration and validate its results. The deformable image registration pipeline used for the validation and test set has been made publicly available to facilitate the participants.

### 4.4 Future Directions

The *SynthRAD2025* dataset provides an excellent foundation for benchmarking existing and emerging sCT generation algorithms even beyond the *SynthRAD2025* Grand Challenge [21], as most parts of the data will stay publicly available. Future research should focus on integrating state-of-the-art sCT generation algorithms into clinical workflows, particularly for online adaptive radiotherapy. This includes evaluating these algorithms' computational efficiency, robustness against outliers and artifacts, and clinical feasibility in real-time treatment scenarios. While the series of SynthRAD datasets and accompanying challenges already covers five anatomical regions (brain, head-and-neck, thorax, abdomen, and pelvis), future iterations could further expand the datasets to include additional regions, such as extremities, special patient populations, e.g., pediatric patients, or extend to other imaging modalities, such as ultrasound and PET. This would further enhance the dataset's applicability to a broader range of clinical scenarios.

## 5 Conclusion

The *SynthRAD2025* dataset is a resource for the radiotherapy research community. It offers a comprehensive and publicly available dataset for synthetic CT generation. This dataset can drive advancements in personalized cancer care by addressing challenges in image synthesis for radiotherapy. Researchers and other dataset users must be mindful of the dataset's limitations and aim to develop robust, generalizable, and clinically feasible algorithms. The release of the validation and test sets after the challenge will further enable the community to validate and refine their approaches.

## Acknowledgments

The *SynthRAD2025* challenge was funded by a grant from "Stiftungen zu Gunsten der Medizinischen Fakultät der Ludwig-Maximilians-Universität München" awarded to Adrian Thummerer to support the computation costs. Adrian Thummerer received funding from a grant from Deutsche Krebshilfe (70114849). None of the centers received compensation for sharing the dataset.